# Defect Passivation and Förster-Type Energy Exchange in $H_2Pc$-TMD Organic-Inorganic Heterostructures

Š. Mandić[1,2], A. Senkić[*1†], N. Vujičić[1]

[1] Centre for Advanced Laser Techniques, Institute of Physics, Bijenička cesta 46, 10 000 Zagreb, Croatia
[2] Faculty of Physics, University of Rijeka, 51 000, Rijeka, Croatia

*Corresponding author: asenkic@ifs.hr

## ABSTRACT

Organic - inorganic heterostructures (HS) combine the strong light absorption and exciton generation capabilities of organic molecules with the unique excitonic properties of layered transition metal dichalcogenides (TMDs), where the interfacial band alignment dictates the optical response. In this work, we investigate the influence of $H_2Pc$ molecules on CVD-grown $MoS_2$ and $WS_2$ monolayers using correlative microscopy techniques - Kelvin probe force microscopy (KPFM), photoluminescence (PL), and Raman spectroscopy. Comprehensive analysis of both electronic and optical properties provides detailed insights into the energy band alignment in these two HS. Despite their similar band alignments, the heterostructures exhibit strikingly different optical signatures. In the case of $H_2Pc/MoS_2$ HS, the effect of defect healing is more pronounced, while for the $H_2Pc/WS_2$ HS, strong indications of Förster energy transfer are observed. These findings highlight the critical role of transition dipole moment in addition to spectral overlap between donor emission and acceptor absorption in the design of optoelectronic devices.

## INTRODUCTION

Two-dimensional (2D) transition metal dichalcogenides (TMDs), such as $MoS_2$ and $WS_2$, have attracted significant attention due to their direct band gaps in the monolayer limit [1,2], strong excitonic effects [2–4], and potential applications in optoelectronics, photovoltaics, and sensors even at room temperatures [5–8]. To exploit the distinctive properties of monolayer TMDs arising from many-body interactions, it is crucial to precisely tune their electro-optical characteristics and transfer these unique features to other materials. A promising strategy involves the formation of heterojunctions, such as TMD/TMD [9–11] and molecule/TMD heterostructures (HS) [12,13]. Although the practical use of TMDs and related HS is often limited by intrinsic point defects and surface states that affect their intrinsic optical, electronic and mechanical properties [14–17], defects offer unique opportunities for functionalization strategies by molecular chemistry approaches, by acting as primary (re)active sites in TDMs [18,19].

One promising approach to enhance the electro-optical performance of TMDs is through non-covalent interactions with conjugated organic molecules [20]. Among these, metal-free phthalocyanine ($H_2Pc$) stands out for its rich electronic structure, strong light absorption, and ability to form ordered van der Waals (vdW) interfaces [21–23]. When deposited onto TMD surfaces, $H_2Pc$ can engage in interfacial interactions that influence charge carrier dynamics [24], and surface potential [25], modulating their electronic and optical properties. It was shown that integrating $H_2Pc$ with monolayer $MoS_2$ accelerates photodetection response by nearly two orders of magnitude without sacrificing responsivity, attributed to improved photocurrent dynamics via intensity-modulated photocurrent spectroscopy [26]. Besides more common process of charge transfer across the interface, under resonant excitation conditions, energy transfer processes like Förster resonance

[†] Current affiliation: Department of physics, University of Münster, Münster, Germany

energy transfer (FRET) is possible [27]. For this process to occur, the donor emission spectrum and the acceptor absorption spectrum must overlap. Previous reports [28,29] have shown high efficiency of FRET from the TMD to the organic layer under resonant excitation conditions. This growing body of work positions $H_2Pc$ as a versatile interfacial modifier, enabling the development of advanced TMD-based devices for optoelectronics [30], flexible electronics, and spintronics [31].

In this study, we investigate the effects of $H_2Pc$ deposition on CVD-grown monolayers of $MoS_2$ and $WS_2$ using a combination of complementary techniques such as atomic force microscopy (AFM), Kelvin probe force microscopy (KPFM) with Raman and photoluminescence (PL) spectroscopy. Unlike the previously published literature, which mainly focuses on molybdenum-based TMDs, this is the first time, at least to our knowledge, that organic-inorganic HS of $H_2Pc$ molecules on monolayer $WS_2$ are being investigated. Considering that CVD samples have an intrinsic concentration of defect states, this research aims to determine how the emission properties of CVD-grown TMDs are modified under the influence of the same organic molecule. Our results show clear signatures of defect healing, exciton modification, and interfacial energy transfer in these organic-inorganic hybrid systems. By comparing the responses of $MoS_2$ and $WS_2$ to functionalization, we also reveal material-specific differences in molecular coupling and doping effects due to the intrinsically different electronic structure of the host layer. This work provides a detailed understanding of organic-inorganic hybrid interfaces, with implications for engineering 2D material-based devices with tailored optoelectronic properties.

## METHODS
### $MoS_2$ synthesis

Monolayer $MoS_2$ was synthesized using a chemical vapor deposition (CVD) method adapted from [32]. A 10 μL droplet of a mixture (1:1 ratio) of precursor solutions containing ammonium heptamolybdate (($NH_4)_6Mo_7O_{24}$, 15.4 ppm in deionized water) and sodium molybdate ($Na_2MoO_4 \cdot 2H_2O$, 15.4 ppm in deionized water) was deposited onto cleaned substrates and dried on a hot plate at 120 °C. The growth was conducted on $Si/SiO_2$ substrates with a 290 nm oxide layer. The substrate was placed in a CVD furnace, where the Mo precursor was thermally decomposed at 650 °C for 30 minutes under argon flow. Sulfur powder was placed in a separate boat and heated independently to 140 °C. The growth temperature was 850°C with argon flow set to 75 sccm. The total growth time was 10 minutes.

### $WS_2$ synthesis

The growth was conducted on $Si/SiO_2$ substrates with a 290 nm oxide layer. For $WS_2$, $H_2WO_4$ (Sigma Aldrich) was diluted first with $NH_3$ in 1:9 ratio. Then DI water was added to obtain 75 ppm solution. This W precursor was mixed with 5 ppm NaOH solution in 1:1 ratio and a 10 μL droplet was deposited onto cleaned substrate and dried on a hot plate at 130 °C. The substrate was placed in a CVD furnace, where the W precursor was thermally decomposed at 500 °C for 10 minutes under argon flow. Sulfur powder was placed in a separate boat and heated independently to 140 °C. The growth temperature was 850°C with argon flow set to 100 sccm. The total growth time was 5 minutes.

### Molecular evaporation

Organic semiconductor phthalocyanine ($H_2Pc$, Alfa Aesar), purified by threefold temperature-gradient sublimation, was deposited in a physical vapor deposition (PVD) chamber (HHV Advanced Technologies, Auto 306), at pressures lower than $3\times10^{-6}$ mbar, under conditions typically yielding a rate of ~ 0.5-1 Ås$^{-1}$. Source shutter was kept open for the duration of 5 seconds.

### AFM and KPFM

AFM and KPFM measurements were performed using a Bruker Nano-Wizard 4 ULTRA AFM system with Pt/Ir-coated conductive cantilevers (NCPt-50, Nano World), featuring a nominal spring constant of 42 N/m and a resonant frequency of 285 kHz. Topography and surface potential images were acquired simultaneously in AC mode. In KPFM mode, the contact potential difference

($V_{CPD}$) was measured by applying a DC bias to nullify the electrostatic force between tip and sample. The sample work function ($\phi_{sample}$) was calculated using [33]:

$$\phi_{sample} = \phi_{probe} - eV_{CPD},$$

where $\phi_{probe}$ is the calibrated work function of the AFM tip. Calibration was performed on freshly cleaved highly ordered pyrolytic graphite (HOPG) prior to each measurement series. From the literature [34] it is known that the work function of HOPG is 4.6 eV. The tip work function was determined to be 4.93 eV for as-grown samples and 4.91 eV for functionalized samples.

**Optical measurements**

Commercial Renishaw in-via Raman setup was used for Raman and photoluminescence (PL) measurements in a back-scattered configuration. The setup is equipped with a 532 nm (2.33 eV) continuous wave laser source and three gratings with 150, 600 and 2400 mm$^{-1}$ constants. For room temperature measurements, the 100x objective with NA = 0.9 was used. Temperature-dependent measurements were conducted with optical flow-cryostat (Cryo Industries of America) and liquid nitrogen as a coolant, with the Olympus LMPlan FLN 50x objective with numerical aperture of 0.5. For Raman (PL) measurements grating with 2400 (150) mm$^{-1}$ constant was used, with laser power on the sample ~ 0.5 mW while the acquisition time was changed, depending on the sample.

A custom-built micro-Raman setup based on a confocal microscope in backscattering geometry, with 488 nm (2.54 eV) excitation was also employed for PL and Raman measurements. A 50× objective (N.A. 0.75) focused the beam to a spot size below 0.85 μm. The backscattered signal was filtered using longpass Raman filter (Semrock Raman RazorEdge Beamsplitter 488 RU) and bandpass filter (Semrock RazorEdge LP Edge Filter 488 LP), then coupled into a 50 μm core fiber serving as a confocal pinhole. The signal was directed to a 50 cm spectrometer (Andor Shamrock 500i-B1) equipped with three gratings (150, 300, and 1800 lines/mm) and detected by a thermoelectrically cooled EMCCD. For PL measurements, a 150 lines/mm grating was used, covering a spectral range of 485-850 nm with an acquisition time of 20x0.05 s. Raman spectra were acquired using the 1800 lines/mm grating with an acquisition time of 20x4 s. The incident laser power at the sample was maintained at 100 μW.

# RESULTS
## Morphological and electronic properties

AFM topography images showed that as-grown MoS$_2$ surfaces exhibit smooth, continuous monolayers characteristic of high-quality CVD growth (Figure 1a). In contrast, as-grown WS$_2$ (Figure 2a) showed nanoscale spherical residues attributed to synthesis-related byproducts [35]. Following H$_2$Pc deposition, both MoS$_2$ (Figure 1b) and WS$_2$ (Figure 2b) exhibited increased surface roughness, with evenly distributed nanoscale aggregates indicating successful physisorption of H$_2$Pc molecules. The root-mean-square (RMS) roughness, calculated from areas indicated by yellow squares on AFM images, increased from (0.20 ± 0.02) nm to (1.02 ± 0.12) nm for MoS$_2$, and from (0.88 ± 0.27) nm to (1.47 ± 0.29) nm for WS$_2$, confirming molecular coverage.

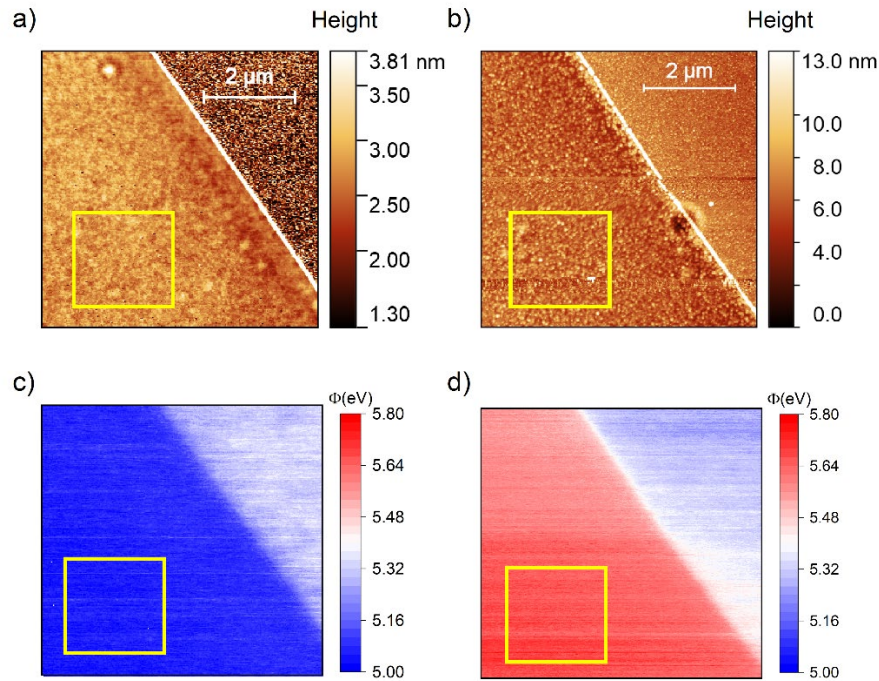

**Figure 1.** AFM topography image of as-grown $MoS_2$ (a) and $H_2Pc/MoS_2$ HS (b). Corresponding KPFM image of as-grown $MoS_2$ (c), and $H_2Pc/MoS_2$ HS (d). Yellow square denotes an area from which RMS roughness values are calculated.

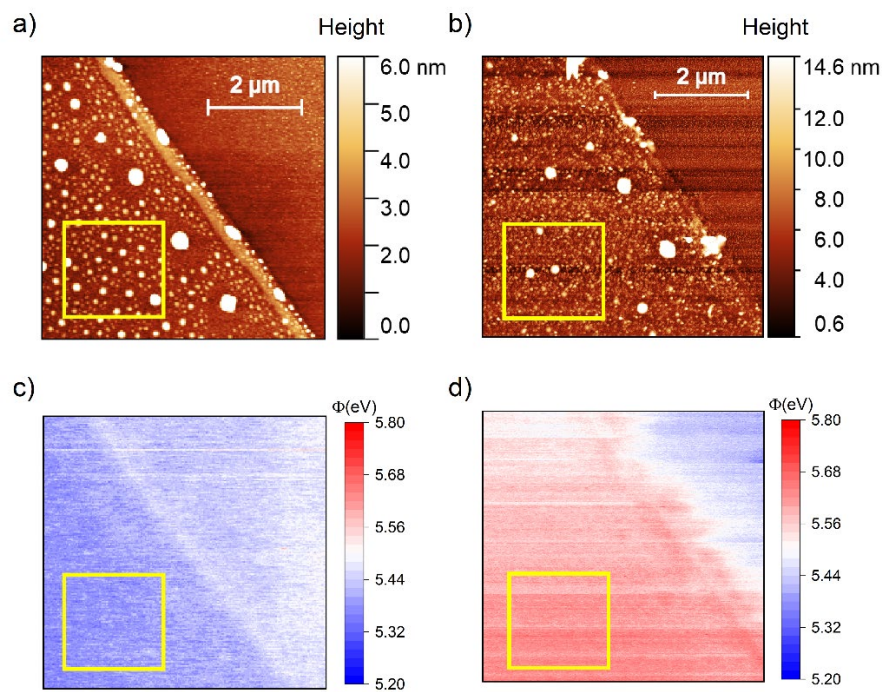

**Figure 2.** (a) AFM image of as grown $WS_2$, (b) AFM image of $H_2Pc/WS_2$ HS, (c) KPFM image of as grown $WS_2$, and (d) KPFM image of $H_2Pc/WS_2$ HS. Yellow square denotes an area from which RMS roughness values are calculated.

KPFM measurements, shown in Figure 1c (2c) for $MoS_2$ ($WS_2$) and Figure 1d (2d) for $H_2Pc/MoS_2$ ($H_2Pc/WS_2$), revealed a substantial increase in work function for both materials following H$_2$Pc deposition, indicating electronic interaction at the molecule/TMD interface. Work function was determined from the same area as the RMS roughness. In MoS$_2$, the work function increased by 0.45 eV, namely from (5.15 ± 0.01) eV to (5.60 ± 0.05) eV. In WS$_2$ the increase was smaller, 0.23 eV, from (5.36 ± 0.01) eV to (5.59 ± 0.02) eV. These increases of work function suggest the formation of interfacial dipoles and charge transfer from the TMDs to the electron-accepting H$_2$Pc molecules, consistent with p-type doping behavior [36]. The spatial distribution of the work function remained uniform across both HSs, further supported by the homogeneous molecular coverage observed in AFM. While both materials exhibit comparable electronic modulation upon H$_2$Pc deposition, the slightly higher absolute work function values for MoS$_2$ may reflect subtle differences in their intrinsic electronic structure and interaction strength with H$_2$Pc, as will be discussed later.

## Optical properties
### Photoluminescence of MoS$_2$ and H$_2$Pc/MoS$_2$ HS

Schematic representation of H$_2$Pc molecule is shown in Figure 3a. Structurally, it consists of a macrocycle with four fused isoindole units surrounding a central cavity where the two hydrogen atoms are located. Isoindole structure consists of a benzene ring fused with a pyrrole-like ring. Four isoindole subunits are linked via nitrogen atoms, forming a square planar H$_2$Pc molecule with resulting D$_{2h}$ symmetry [37]. To investigate excitonic behavior and emission characteristics of the TMD layer and related heterostructures, PL measurements were performed on monolayers before and after molecular deposition, and on H$_2$Pc molecules deposited directly on SiO$_2$/Si substrate. Optical properties of organic molecules depend on substrate, thickness, and aggregation type [38]. Figure 3b shows emission spectra of H$_2$Pc molecules deposited on the substrate measured at 77 K and 290K. Emission at 1.76 eV is attributed to the zero-phonon transition [39], called Q band, and does not show significant shift with temperature increase. When H$_2$Pc is in solid phase, its excitonic coupling usually causes the formation of physical dimers, which allows the excited state to oscillate between two molecules [40], explaining why zero phonon transition has smaller intensity and broader emission line. Emissions between 1.2 and 1.55 eV are characteristic for H$_2$Pc aggregation in samples thicker than 1 nm or in case of single crystals [39,41]. The emission spectra exhibit a cut-off near 1.9 eV, attributed to an absorption edge in the 2.07-2.10 eV range, which defines the optical bandgap corresponding to the HOMO-LUMO transition [42].

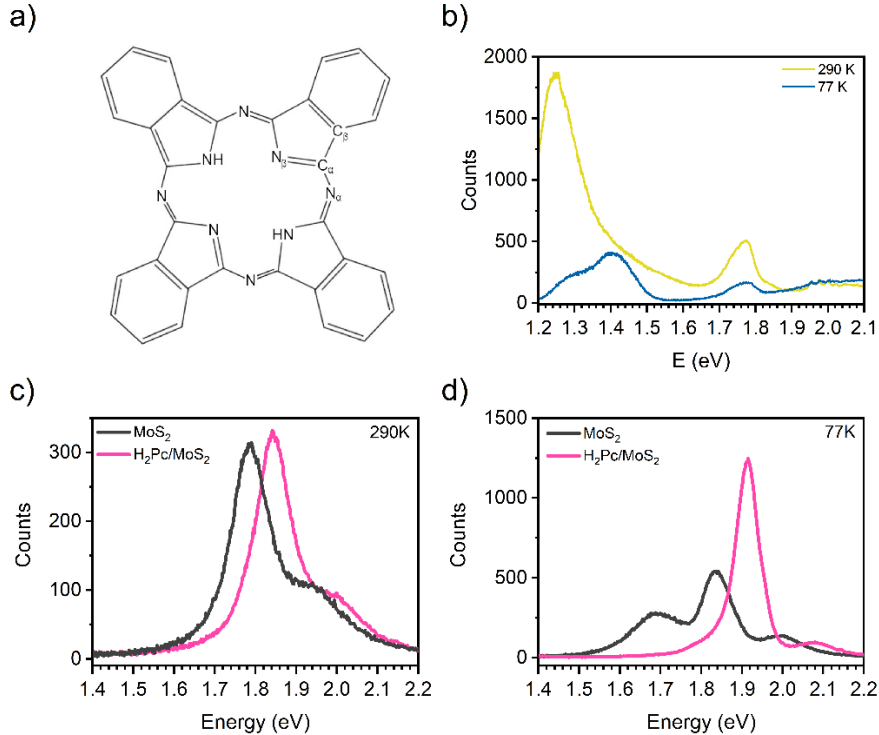

**Figure 3.** (a) Schematic representation of H$_2$Pc molecule. (b) Emission spectra of H$_2$Pc molecules deposited on SiO$_2$/Si substrate taken at 77 K (blue) and 290 K (yellow). (c) PL spectra of as-grown MoS$_2$ (black) and H$_2$Pc/MoS$_2$ (pink) samples on SiO$_2$/Si substrate at 290 K and (d) 77 K. The PL spectra were obtained by exciting the samples with 2.33 eV laser.

The optical spectra of TMDs are characterized by two prominent excitonic features, labeled A and B excitons [43]. The optical bandgap of as-grown MoS$_2$ is known from our previous work [44] and is related to the emission energy of the A exciton. Considering the bandgap energies and the associated band structure [24,45], hole transfer from the HOMO level to the MoS$_2$ valence band is energetically favorable at the interface. The PL spectra of MoS$_2$ and the H$_2$Pc/MoS$_2$ heterostructure at room temperature (290 K) and 77 K are shown in Figures 3c and 3d, respectively. At room temperature, the PL intensity of the H$_2$Pc/MoS$_2$ heterostructure slightly increases compared to as-grown MoS$_2$, accompanied by a blueshift of the A and B exciton energies [46], consistent with p-type doping. At 77 K, the PL spectra of as-grown MoS$_2$ and the H$_2$Pc/MoS$_2$ HS show noticeable differences. The spectrum of as-grown MoS$_2$ exhibits an additional peak at (1.684 ± 0.001) eV, attributed to a bound exciton, which is absent in the H$_2$Pc/MoS$_2$ HS. These defect-bound excitons ($X_B$) are associated with localized in-gap states in which free excitons become trapped, leading to variations in PL emission intensity and energy. This in-gap luminescence becomes more pronounced with an increasing number of defects and correlates with the spatial distribution of defects inherent to CVD-grown materials [17]. The PL spectrum of the H$_2$Pc/MoS$_2$ HS at 77 K is blue-shifted, with a narrower A exciton spectral line, indicating longer exciton lifetimes. Optical maps presented in Figures S1 and S2 in the Supplementary Information (SI) show that this behavior is present throughout the entire sample, indicating significantly improved quality. The intensity ratio of A and B excitons can be used to qualitatively assess non-radiative recombination, where a low B/A ratio indicates low defect density and high sample quality [47]. B/A ratio decreases from 0.27 to 0.19 after molecular deposition. These results suggest that H$_2$Pc efficiently heals intrinsic defects in CVD grown MoS$_2$.

## Photoluminescence of $WS_2$ and $H_2Pc/WS_2$ HS

In the case of $WS_2$, PL spectra revealed a rich excitonic structure following $H_2Pc$ deposition. At the temperature of 290 K (Figure 4a) the as-grown $WS_2$ sample displayed a dominant A exciton peak at $(1.9652 \pm 3·10^{-4})$ eV. Upon the deposition of $H_2Pc$ molecules, this emission undergoes a blueshift to $(1.9693 \pm 3·10^{-4})$ eV [46], confirming KPFM measurements that $H_2Pc$ acts as a p-type dopant. Another notable difference is that the A exciton intensity is substantially reduced in the $H_2Pc/WS_2$ HS.

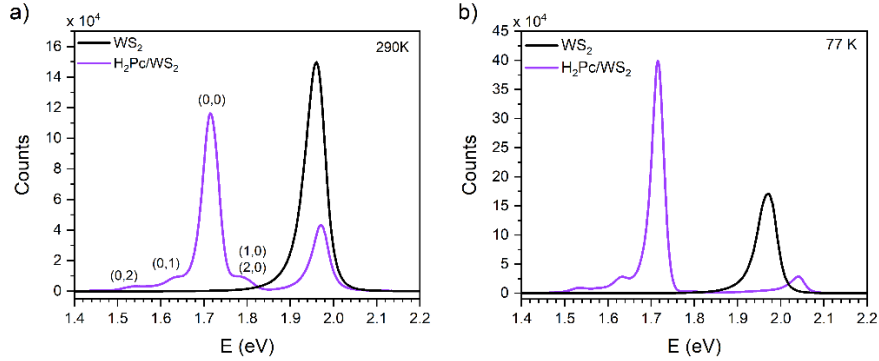

**Figure 4.** PL spectra of as-grown $WS_2$ and $WS_2/H_2Pc$ HS (a) at 290 K, and (b) at 77 K.

In the spectral region of 1.50 to 1.80 eV, new emissions arising from the $H_2Pc$ molecule are present. They are in good accordance with single-molecule emission reported in [48,49], where the $H_2Pc$ molecule was placed on NaCl(110) spacer on Ag(111) substrate and investigated at 4K using scanning tunneling microscope. In cited papers, the emission line is significantly narrower, because it stems from a single molecule, and not thin film, where intermolecular interactions broaden the electronic features.

In our case, the most intense emission observed at $(1.7149 \pm 2·10^{-4})$ eV, is attributed to zero-phonon (0-0) line, corresponding to the lower $Q_x$ state. On each side of the zero-phonon emission there are two vibronic satellite peaks, (2-0) at $(1.806 \pm 0.003)$ eV, (1-0) at $(1.786 \pm 0.004)$ eV, (0-1) at $(1.6293 \pm 6·10^{-4})$ eV, and (0-2) at $(1.5369 \pm 5·10^{-4})$ eV. The sidebands in our system are quite broadened with comparison to the ones reported in [49], which is expected since: 1) the spectra were recorded on a thin $H_2Pc$ layer and not a single molecule, 2) the spectra were recorded at higher temperatures and 3) thin $H_2Pc$ layer acts as a p-dopant to the $WS_2$ monolayer underneath. The change in the emission line-shape, accompanied with the red shift of 20 meV compared to the $H_2Pc$ on $SiO_2/Si$ is a direct consequence of strong interaction between $H_2Pc$ and $WS_2$. The emergence of molecular vibronic states alongside the quenching of the A exciton PL signal suggests a FRET mechanism, in which $WS_2$ acts as the energy donor and $H_2Pc$ as the acceptor [50].

The efficiency of FRET process depends on the spectral overlap of the donor emission and the acceptor absorption spectrum [51]. The $H_2Pc$ Q-band emission is in the 1.7 - 1.8 eV window, as shown in Figure 3a, but it is important to mention that its absorption spectrum is blueshifted [52]. Moreover, the efficiency non-trivially depends on the Förster distance d: for small distances the efficiency varies as $e^{-d}$, while at larger distances it changes to $d^{-4}$ [53,54]. Lastly, the efficiency is proportional to the squares of transition dipole moments of both constituents in HS [55]. Since the monolayer $WS_2$ has larger transition dipole moment than $MoS_2$ [56], dipole-dipole interaction is stronger in $H_2Pc/WS_2$ HS, making the FRET process more efficient in $H_2Pc/WS_2$ HS, even though $WS_2$ emission spectrum has narrower spectral overlap with the absorption spectrum of $H_2Pc$. Since $H_2Pc$ has a resonant Raman process with 2.33 eV laser excitation [52,57], we also conducted off-resonant Raman and PL measurements, using home-made PL/Raman microscope with 2.48 eV laser. The PL spectra of both HSs are shown in Figure S3 in SI. In $H_2Pc/WS_2$ HS, the PL intensity of $H_2Pc$ molecules is smaller than of $WS_2$, due to the prolonged environmental exposure. Additionally, smaller pinhole size in the home-made setup, coupled with longer spectrometer size, makes the detection of near-infrared photons less efficient than in the commercial setup. Nevertheless, since there are no changes in the emission line shapes in both HSs, this supports our conclusion that there is a significant energy transfer from $WS_2$ to $H_2Pc$ molecules, due to a much larger transition dipole moment of monolayer $WS_2$ with respect to $MoS_2$.

Figure 4b shows PL spectra of $WS_2$ and $H_2Pc/WS_2$ HS at 77K. Regarding the $WS_2$ emission in $H_2Pc/WS_2$ HS, apart from the blueshift of A exciton line due to presence of $H_2Pc$ molecules, its decreased intensity reveals another emission at $(1.94 \pm 0.01)$ eV, stemming from the $X_B$ excitons, which is confirmed with power-dependent measurements at 77 K (Figure S4 in SI). The $X_B$ contribution at 77K in as-grown $WS_2$ is not observed due to large intensity of A exciton emission.

These can be attributed to the strong interaction between the $H_2Pc$ layer and $WS_2$ at the interface. The pronounced difference in electron-phonon coupling strength compared to $H_2Pc/MoS_2$ arises from variations in the interaction strength between monolayer TMDs and the molecules, as well as from differences in their specific band structures and transition dipole moments [56]. Regarding the temperature dependence of A exciton's FWHM, in case of as-grown $WS_2$, it does not significantly vary with the increase of temperature, while slight linear increase is observed in case of $H_2Pc/WS_2$ HS, supporting the evidence of strong vibronic coupling, since the inhomogeneous broadening is the most likely cause of this FWHM increase with temperature.

## Temperature dependent PL measurement

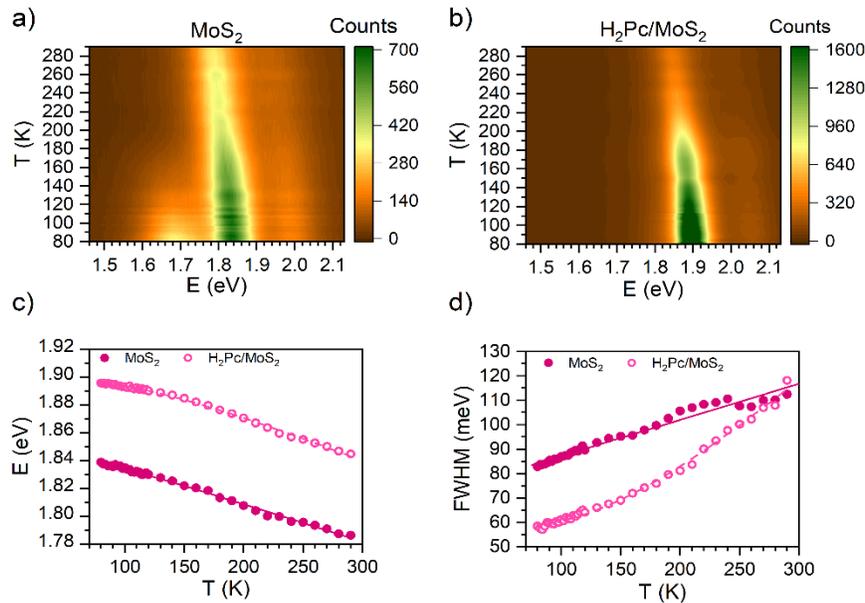

**Figure 5.** (a) Temperature dependent PL spectra of (a) as-grown $MoS_2$ and (b) $MoS_2/H_2Pc$ HS measured in the temperature range from 77 up to 290 K. Temperature dependence of A exciton (c) energy and (d) FWHM for both samples. Lines in the d) panel are eye-guides.

Figures 5a and 5b show the PL spectra evolution of as-grown $MoS_2$ and $H_2Pc/MoS_2$ HS measured in the temperature range from 77 K up to 290 K. Low-temperature PL spectrum of as-grown $MoS_2$ is broader and redshifted compared to emission spectra of $H_2Pc/MoS_2$ HS. Reduction in the ratio of B- and A-exciton indicates improvement in sample quality as in Figure 5b B-exciton has reduced intensity. The PL emission of $X_B$ in as-grown $MoS_2$ quenches with increasing temperature and becomes unnoticeable at 190 K. Photoluminescence spectra of as-grown $MoS_2$ ($H_2Pc/MoS_2$ HS) are fitted as a sum of three (two) Lorentzian functions from which information about energy, FWHM and intensity of emission lines were extracted. For the same temperature range, the redshift of the as-grown $MoS_2$ is smaller (45.4 meV) than in the case of $H_2Pc/MoS_2$ HS

(50.8 meV), as presented in Figure 5c. The larger thermal band shift observed for the HS is directly associated to the influence of phonons on electronic lines, i.e., strong sign of vibronic coupling in the system [41]. Figure 5d shows decrease in FWHM after molecular deposition indicating reduced disorder and decrease in phonon scattering at 77 K. Increase in temperature causes change in interfacial coupling in HS and stronger intermolecular interactions affecting charge transfer dynamics and electron-phonon scattering leading to almost equal values FWHM at room temperature.

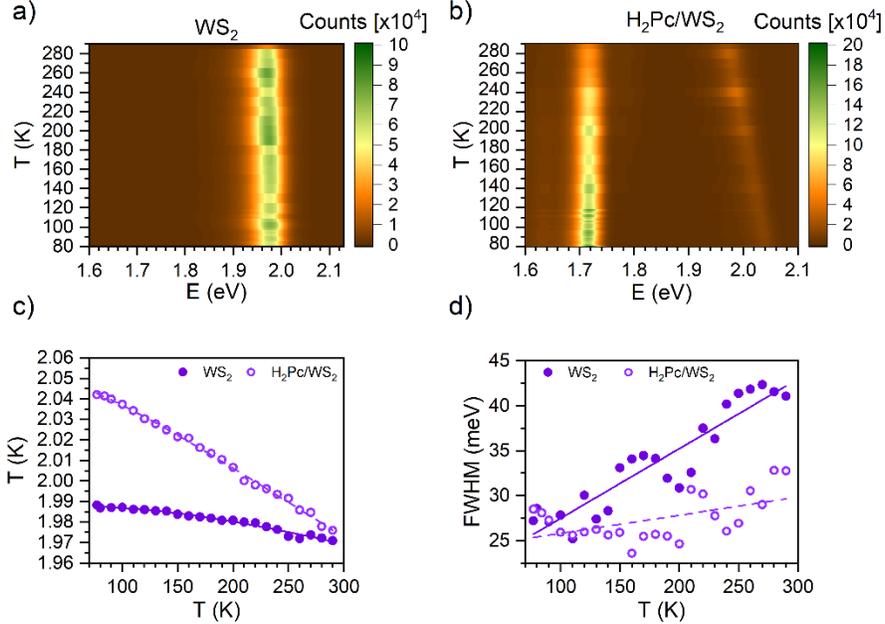

**Figure 6.** Temperature dependent PL spectra of (a) as-grown $WS_2$ and (b) $WS_2/H_2Pc$ HS measured in the temperature range from 77 up to 290 K. Temperature dependence of A exciton (c) energy and (d) FWHM for both samples. Lines in the d) panel are eye-guides.

Temperature dependence of PL spectra for as-grown $WS_2$ and the $H_2Pc/WS_2$ HS is presented in Figures 6a and 6b, respectively. A notable difference is observed in the redshift of the A-exciton between the samples with the temperature increase (Figure 6c). Specifically, the redshift in as-grown $WS_2$ is smaller (17.3 meV) compared to that in the $H_2Pc/WS_2$ HS (66.5 meV), with both converging to nearly the same energy at room temperature. This can be attributed to enhanced electron-phonon coupling in the HS, a topic that will be addressed in more detail later. In addition to the redshift, a temperature-induced suppression of FRET is evident, as the A exciton PL intensity increases while the intensity of molecular zero-phonon line decreases with rising temperature. Furthermore, the temperature dependence of the A-exciton FWHM, shown in Figure 6d, reveals a slower increase in the heterostructure compared to the as-grown material. This indicates a reduced rise in disorder, possibly mediated by temperature-dependent energy and charge transfer dynamics.

**Table 1.** List of the fit parameters for modified Varshni relation.

| Material | $E_0$ (eV) | S (-) | $E_{ph}$ (meV) | Phonon |
|---|---|---|---|---|
| $MoS_2$ | 1.846 ± 0.002 | 1.61 ± 0.06 | 13 ± 3 | TA(Q) [58] |
| $H_2Pc/MoS_2$ | 1.8972 ± 0.0004 | 2.11 ± 0.07 | 30 ± 1 | LA(K) [58] |
| $WS_2$ | 1.9876 ± 0.0005 | 0.8 ± 0.1 | 39.1 ± 0.5 | TO(M) [58] |
| $H_2Pc/WS_2$ | 2.049 ± 0.002 | 2.04 ± 0.08 | 16 ± 3 | LA(M) [58] |

The energy band gap of semiconducting materials redshifts with an increase in temperature due to electron-phonon interactions. The phonon energy can be extracted after fitting temperature-dependence of the emission energy to modified Varshni equation:

$$E(T) = E_0 - S \cdot E_{ph}[\coth\left(\frac{E_{ph}}{2k_B T}\right) - 1], \quad (1)$$

where $E_0$ is the energy bandgap at 0 K, $E_{ph}$ is the phonon energy, $k_B$ is Boltzmann's constant, T is temperature and S is the Huang-Rhys factor - a measure of the electron-phonon coupling strength [42]. Table 1 summarizes the values for all parameters in Equation (1).

For $MoS_2$, the phonon energy of 12 meV corresponds to the transverse acoustic (TA(Q)) mode. Such low-energy phonons are liked to enhanced scattering at defect sites. In case of $H_2Pc/MoS_2$ HS, the phonon energy increases to 30 meV, reaching values obtained in non-aged CVD-grown samples [44], confirming the defect passivation process in aged CVD-grown samples. Both modes are known to be the dominant acoustic phonon excitations responsible for significant electron-phonon scattering in monolayer $MoS_2$ at low temperatures [43]. The influence of acoustic phonons on excitonic emission lines is further supported by the temperature-dependent broadening of the spectral lines, which shows a clear linear behavior (see Figure 5d) [59]. The relatively small increase in electron-phonon coupling from 1.61 in as-grown material to 2.11 in the HS accompanied with the acoustic phonons stiffening is consistent with hole doping [60].

In the case of $WS_2$ and its corresponding HS, molecular deposition leads to a significant change in the Huang-Rhys factor. The strength of electron-phonon coupling more than doubles, indicating enhanced charge interactions [61], which become abundant following molecular deposition. These processes in the HS further support the presence of interfacial dipoles and charge transfer across the interface [62]. Additionally, this phenomenon is accompanied by a softening of acoustic phonon modes, attributed to charge doping effects [63].

**Raman spectroscopy**

To investigate the influence of $H_2Pc$ molecules on the vibrational spectra in TMDs, we conducted room-temperature Raman measurements.

As-grown $MoS_2$ monolayer has two intralayer vibrational modes (Figure 7a), corresponding to the in-plane $E_{2g}$ and out-of-plane $A_{1g}$ phonon modes, with frequencies of (382.98 ± 0.04) cm$^{-1}$ and (402.49 ± 0.06) cm$^{-1}$, respectively [64–66]. Upon $H_2Pc$ deposition, frequencies of these two Raman modes slightly increase, confirming the KPFM measurements that $H_2Pc$ acts as a p-dopant for monolayer $MoS_2$ [67] (Table S2 in SI).

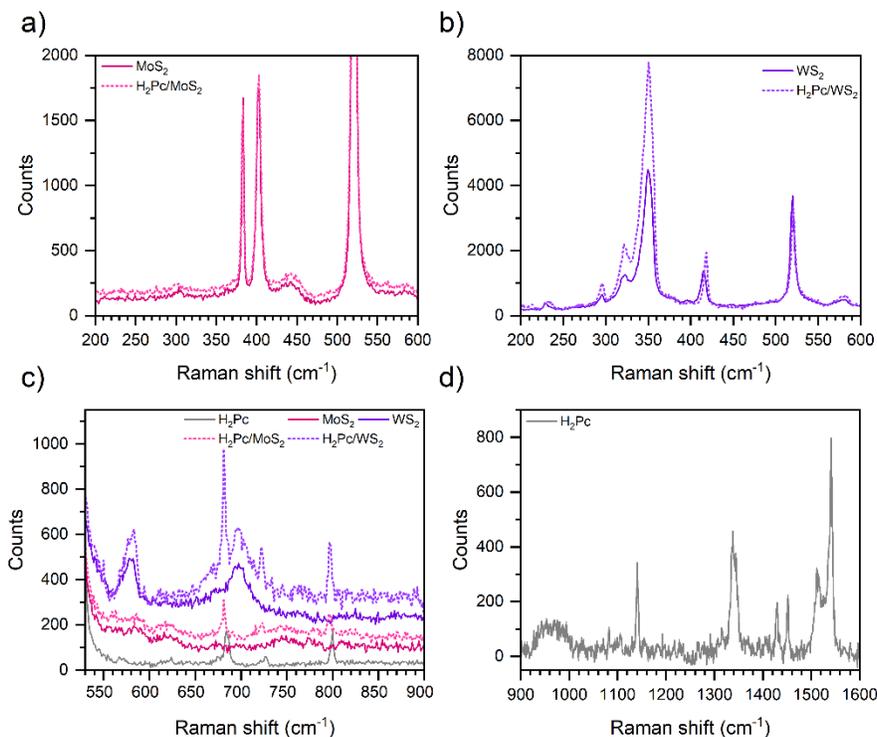

**Figure 7.** Raman spectra of as-grown MoS$_2$ and H$_2$Pc/MoS$_2$ HS (a), and WS$_2$ and H$_2$Pc/WS$_2$ HS (b), all samples and H$_2$Pc on SiO$_2$/Si substrates (c) and H$_2$Pc on SiO$_2$/Si in the fingerprint region (d). Spectra have small vertical offset for clarity.

Regarding the as-grown WS$_2$ sample (Figure 7b), higher-order Raman modes are also visible due to the resonance effect of transition in tungsten atom with 2.33 eV excitation laser [68]. The in-plane E$^1_{2g}$(Γ) and the out-of-plane A$_{1g}$(Γ) modes have frequencies of (354.6 ± 0.2) cm$^{-1}$ and (416.02 ± 0.08) cm$^{-1}$, respectively [69]. The frequencies of other modes are listed in the Table S2 in SI. Following H$_2$Pc deposition, the most significant change occurs in the stiffening of the A$_{1g}$(Γ) mode, by 2 cm$^{-1}$, confirming significant p-doping.

Changes in the frequencies of molecular Raman modes indicate modifications in the chemical bonds. A detailed assignment of modes in 550 - 1600 cm$^{-1}$ range for H$_2$Pc on SiO$_2$/Si substrate is provided in Table S2 in SI, while spatial maps of all important H$_2$Pc Raman modes are shown in Figures S5 in SI. It is known from literature [70] that excitons in π-conjugated molecules are strongly coupled to the C-C stretching modes with frequencies above 1000 cm$^{-1}$. But, as shown in Figures S5 in SI, the most striking difference occurs for 1340, 1450 and 1540 cm$^{-1}$ modes, related to the deformations of the Pc macrocycle. In the case of H$_2$Pc/WS$_2$ HS, frequencies of these three modes increase by ∼ 2 cm$^{-1}$, with respect to H$_2$Pc deposited on the SiO$_2$/Si substrate, while for H$_2$Pc/MoS$_2$ HS the effect is spatially inhomogeneous and not so pronounced. Pc macrocycle deformations at 670 and 1340 cm$^{-1}$ are associated with Q-band excitation [71] - indicating that the Q-band is efficiently excited with underlying WS$_2$, supporting the conclusion that in H$_2$Pc/WS$_2$ HS, energy transfer is the dominant exchange process. On the other hand, frequency of the mode at 1140 cm$^{-1}$ increases in both HSs, but the effect is more pronounced in H$_2$Pc/MoS$_2$ HS. Unlike the previous two modes which are coupled with the intramolecular electronic transitions, this mode is associated with the lattice vibration caused by the charge transfer between two neighboring molecules [71], indicating larger concentration or agglomeration of H$_2$Pc molecules on MoS$_2$ sample.

Based on these results, we propose two competing mechanisms that govern the processes in these functional HSs: (i) charge transfer from the TMD layer to the molecular layer, and (ii)

FRET (Figure 8.). Charge transfer is evident in optical properties of both HSs and further confirmed by complementary KPFM measurements. Due to the suppression of PL emission related to defect-bound states in $H_2Pc/MoS_2$ HS, which is accompanied by a lack of PL emission of $H_2Pc$, we conclude that charge transfer is the dominant process in $H_2Pc/MoS_2$ HS. On the other hand, in the case of $H_2Pc/WS_2$ HS, efficiency of FRET process is enhanced because of two factors: 1) monolayer $WS_2$ has significantly higher transition dipole moment than monolayer $MoS_2$, 2) the Förster distance between $H_2Pc$ and underlying TMD material is smaller for $WS_2$, as calculated from the AFM roughness data. Strong interfacial coupling is evidenced by the increased intensity of molecular Raman modes on the TMD layer compared to the $SiO_2/Si$ substrate, an effect known as surface enhanced Raman scattering (SERS).

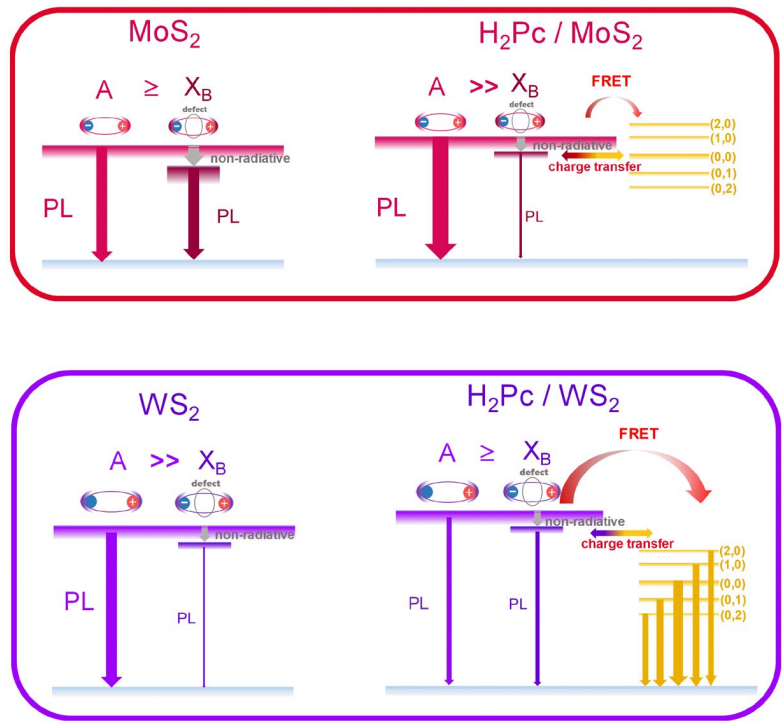

**Figure 8.** Charge and Förster resonant energy transfers (FRET) in $H_2Pc$/TMD heterostructures (HS). The intensity of photoluminescence (PL) emission lines is denoted with the arrow width – increased PL intensity corresponds to a thicker arrow. Molecular energy levels are denoted with yellow horizontal lines. a) In $H_2Pc/MoS_2$ HS the emission of A exciton from $MoS_2$ is enhanced, while the emission from defect-bound states is suppressed, due to an efficient charge transfer. b) In $H_2Pc/WS_2$ HS the emission of A exciton from $WS_2$ is suppressed, while emission from $H_2Pc$ molecule is observed, due to an efficient resonant energy transfer.

## CONCLUSION

In this study, we examined the effects of $H_2Pc$ molecule deposition on CVD-grown monolayers of $MoS_2$ and $WS_2$ using correlative topographical and optical microscopy techniques. Results show that there are two competing processes in both heterostructures (HS): i) charge transfer from the underlying TMD material to the molecular layer and ii) Förster resonant energy transfer (FRET). Charge transfer was evidenced by the Kelvin probe force microscopy (KPFM) measurements: changes in the work function and Fermi level indicate p-type doping of the TMD layer. Photoluminescence (PL) studies showed, apart from the increase in bandgap energy for both HSs, distinct material-dependent optical responses: in $MoS_2$, defect-bound exciton emission was suppressed, suggesting effective defect passivation, and A exciton emission was significantly narrowed, indicating prolonged exciton lifetimes. These results suggest that the charge transfer is the dominant process in $H_2Pc/MoS_2$ HS. On the other hand, in $H_2Pc/WS_2$ HS, due to large transition dipole moment and smaller distance between $H_2Pc$ and underlying $WS_2$, FRET process is more efficient, as evidenced by quenching of the A exciton and the emergence of the $H_2Pc$ emission. Lastly, analysis of $H_2Pc$ Raman spectra showed that the modes which strongly couple to the electronic states in $H_2Pc$ are significantly stiffened in case of $H_2Pc/WS_2$ HS, further evidencing efficient resonant energy transfer. Overall, our results provide a comprehensive understanding of organic molecule-TMD hybrid interfaces and demonstrate that $H_2Pc$ is an effective functionalization agent for modulating the optoelectronic properties of 2D semiconductors. Our further studies will involve the effects of $H_2Pc$ molecules on pristine, exfoliated TMD samples.

## Author contributions

Šimun Mandić (https://orcid.org/0009-0003-5865-3370): Conceptualization, Methodology, Investigation and Data Analysis- synthesis, PL measurements, KPFM measurements, Systematization, Visualization, Writing - Original Draft; Ana Senkić (orcid.org/0000-0002-0567-5299): Conceptualization, Methodology, Supervision, Investigation and Data Analysis- synthesis, PL measurements, Systematization, Visualization, Writing - Original Draft; Nataša Vujičić (orcid.org/0000-0002-5437-5786): Supervision, Methodology, Conceptualization, Funding acquisition, Writing - Original Draft.

## ACKNOWLEDGMENTS

This work was supported by the project Centre for Advanced Laser Techniques (CALT), co-funded by the European Union through the European Regional Development Fund under the Competitiveness and Cohesion Operational Programme (grant no. KK.01.1.1.05.0001) and a project funded by Croatian Science Foundation, Grant No. UIP-2020-02-8891. Authors also acknowledge financial support from the project "Podizanje znanstvene izvrsnosti Centra za napredne laserske tehnike (CALTboost)", financed by the European Union through the National Recovery and Resilience Plan 2021-2026 (NRRP). Authors gratefully acknowledge prof. dr.sc. Vedran Đerek (Department of Physics, Faculty of Science, University of Zagreb, Zagreb, Croatia) for providing access to the physical vapor deposition chamber for evaporation of molecules. Authors also acknowledge dr. sc. Borna Radatović and mag. chem. Ana Jurković for fruitful discussions.

# Defect Passivation and Förster-Type Energy Exchange in H₂Pc-TMD Organic-Inorganic Heterostructures

## Supplementary information

Š. Mandić[1,2], A. Senkić[*1†], N. Vujičić[1]

[1] Centre for Advanced Laser Techniques, Institute of Physics, Bijenička cesta 46, HR-10000 Zagreb, Croatia
[2] Faculty of Physics, University of Rijeka, 51 000, Rijeka, Croatia

*Corresponding author: asenkic@ifs.hr

Spatially resolved photoluminescence (PL) maps of as-grown $MoS_2$ and $H_2Pc/MoS_2$ HS are taken at 290 K (Figure S1) and 77 K (Figure S2). Note that the colorbars are in nm, not eV like in the main text.

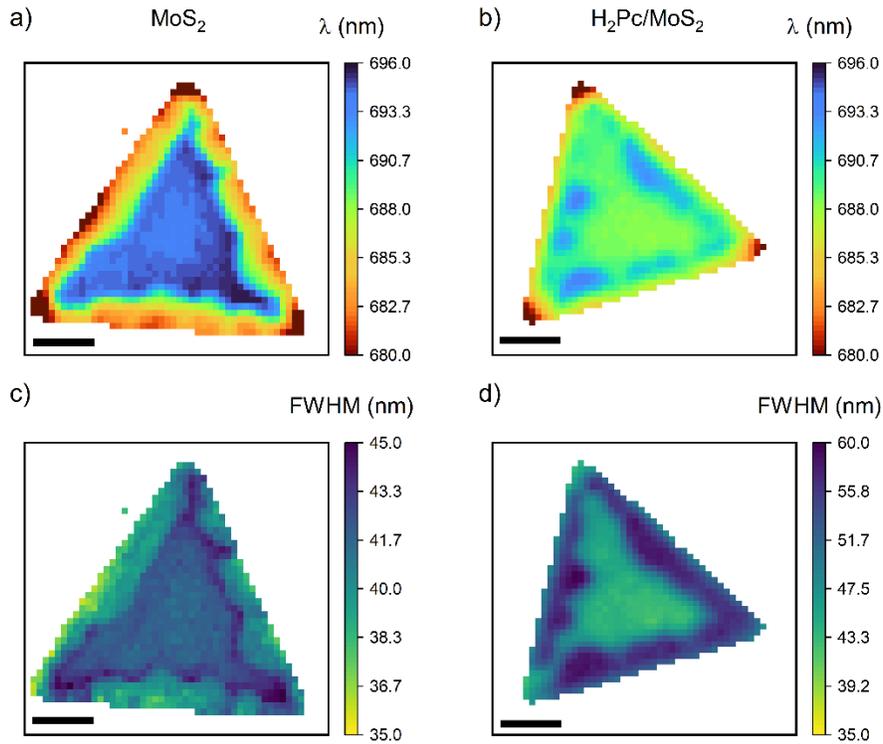

**Figure S1.** Spatially resolved PL maps at 290 K of A exciton wavelength in as-grown $MoS_2$ (a) and $H_2Pc/MoS_2$ HS (b) and its corresponding full width at half maximum (FWHM) (c) and (d). Scalebar is 5µm.

---

[†] Current affiliation: Department of physics, University of Münster, Münster, Germany

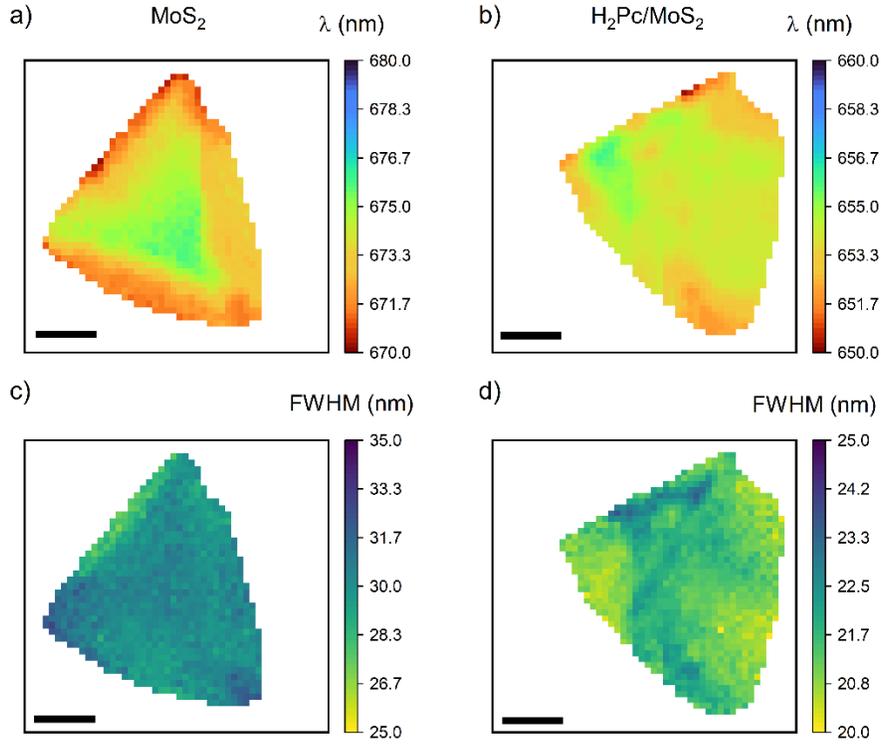

**Figure S2.** Spatially resolved PL maps at 77 K of A exciton wavelength in as-grown $MoS_2$ (a) and $H_2Pc/MoS_2$ HS (b) and its corresponding full width at half maximum (FWHM) (c) and (d). Scalebar is 5μm.

Power-dependent PL measurements at 77 K were analyzed in a usual way [1]. Every spectrum is fitted to a sum of Lorenzian functions and emission's intensity (area under the curve) was determined. The intensity of A exciton emission depends linearly on the excitation power and in general, the dependence of any emission line's intensity on the excitation power can be written as: $I(P) \sim P^{\alpha}$. For semiconductors if the coefficient α is smaller than 1, this emission line stems from defect-bound states, since its intensity saturates with the increase of the laser power. Values of ≈ 1.5 indicate existence of trions, ≈ 2 of biexcitons etc. In Table S1 all fit coefficients are listed for the $H_2Pc/WS_2$ HS.

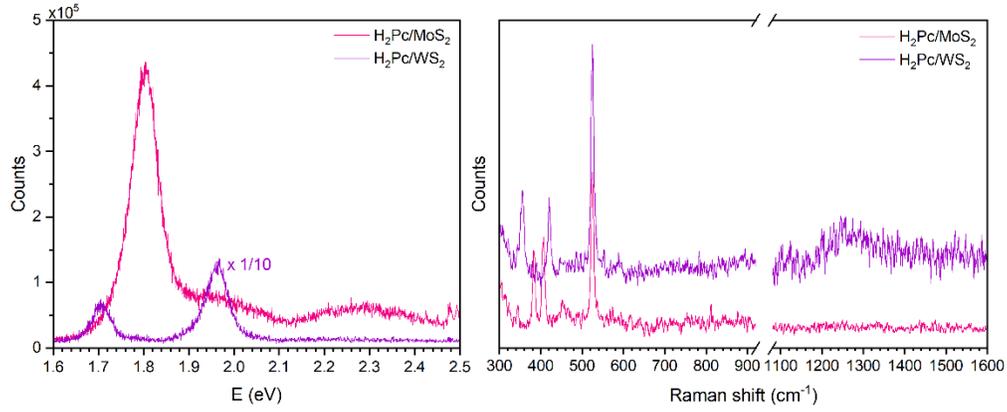

**Figure S3.** Photoluminescence (a) and Raman (b) spectra for $H_2Pc/MoS_2$ and $H_2Pc/WS_2$ HSs with 2.48 eV excitation taken at room temperature.

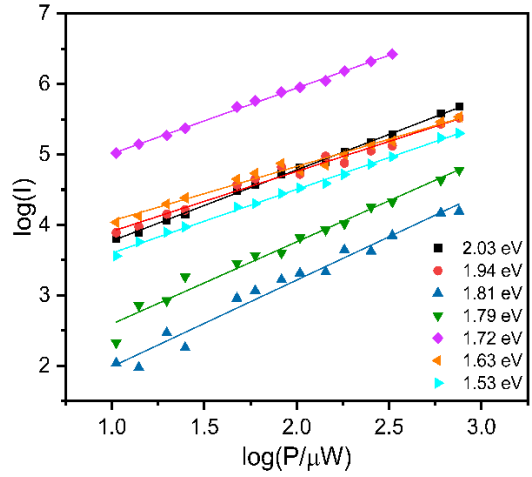

**Figure S4.** Power-dependent measurements of PL intensities for each emission line in the $H_2Pc/WS_2$ HS spectra at 77 K.

**Table S1.** Power-law fit coefficients for all PL components in $H_2Pc/WS_2$ HS at 77K.

| Energy (eV) | Notation | Power-law fit coefficient |
|---|---|---|
| 1.53 | $H_2Pc$ (0-2) | 0.90 ± 0.01 |
| 1.63 | $H_2Pc$ (0-1) | 0.77 ± 0.03 |
| 1.72 | $H_2Pc$ (0-0) | 0.93 ± 0.01 |
| 1.79 | $H_2Pc$ (1-0) | 1.17 ± 0.05 |
| 1.81 | $H_2Pc$ (2-0) | 1.23 ± 0.06 |
| 1.94 | $WS_2$ $X_B$ | 0.85 ± 0.03 |
| 2.03 | $WS_2$ A exciton | 1.01 ± 0.01 |

In Table S2 frequencies and FWHM of all Raman modes are listed for both as-grown samples and for both HS. In the right-most column are listed references from which modes' assignations were taken.

Table S2. List of the fit parameters for Lorenzian functions of Raman spectra.

| Material | Frequency (cm$^{-1}$) | FWHM (cm$^{-1}$) | Assignation | References |
|---|---|---|---|---|
| H$_2$Pc | 569.1 ± 0.9 | 7 ± 3 | Benzene ring deformation | [2,3] |
| | 684.3 ± 0.1 | 5.3 ± 0.3 | Totally symmetric Pc macrocycle breathing mode | [2,3] |
| | 726.7 ± 0.2 | 2 ± 1 | Pc macrocycle deformation mode | [2,3] |
| | 799.63 ± 0.08 | 3.0 ± 0.2 | | |
| | 1140.8 ± 0.2 | 4.4 ± 0.5 | Pyrrole stretching, H-C-C | [4] |
| | 1339.8 ± 0.2 | 14.1 ± 0.7 | Pyrrole stretching In plane full symmetric N-C stretch and ring C-C stretch | [5] |
| | 1429.2 ± 0.3 | 4.2 ± 0.8 | Benzene stretching (C-N$_\alpha$, C-C pyrrole, H-C-C) + (C-N$_\alpha$, C-N$_\alpha$-C, H-C-C) | [4] |
| | 1451.6 ± 0.2 | 3.3 ± 0.6 | 726 cm$^{-1}$ overtone, C-N$_\alpha$-C, H-C-C | |
| | 1515.4 ± 0.4 | 19 ± 1 | C-C pyrrole + benzene, N$_\alpha$, C-C pyrrole, C$_\beta$-C$_\beta$, H-C-C | |
| | 1540.8 ± 0.1 | 7.1 ± 0.3 | Pyrrole stretching, ring C-C stretch and in plane ring symmetric nonmetal bound N-C stretch | [5] |
| MoS$_2$ | 382.98 ± 0.04 | 2.5 ± 0.1 | E$_{2g}$ | [6–8] |
| | 402.49 ± 0.06 | 5.5 ± 0.2 | A$_{1g}$ | |
| H$_2$Pc/MoS$_2$ | 384.19 ± 0.05 | 2.7 ± 0.2 | E$_{2g}$ | |
| | 402.74 ± 0.07 | 6.1 ± 0.2 | A$_{1g}$ | |
| WS$_2$ | 295.7 ± 0.2 | 6.4 ± 0.6 | 2ZA(M) | [9,10] |
| | 322.0 ± 0.4 | 16 ± 1 | TO(M) | |
| | 344 ± 1 | 14 ± 2 | E$^1_{2g}$(M) | |
| | 349.3 ± 0.2 | 8 ± 1 | 2LA(M) | |
| | 354.6 ± 0.2 | 4.7 ± 0.6 | E$^1_{2g}$(Γ) | |
| | 416.02 ± 0.08 | 7.0 ± 0.4 | A$_{1g}$(Γ) | |
| | 579.2 ± 0.3 | 12 ± 1 | A$_{1g}$(M)+LA(M) | |
| | 695.0 ± 0.4 | 34 ± 1 | 4LA(M) | |
| H$_2$Pc/ WS$_2$ | 296.1 ± 0.3 | 6 ± 1 | 2ZA(M) | |
| | 322.9 ± 0.4 | 12 ± 1 | TO(M) | |
| | 343.3 ± 1.4 | 14 ± 3 | E$^1_{2g}$(M) | |
| | 350.8 ± 0.2 | 7.4 ± 0.9 | 2LA(M) | |
| | 355.4 ± 0.1 | 3.7 ± 0.7 | E$^1_{2g}$(Γ) | |
| | 418.08 ± 0.08 | 4.1 ± 0.3 | A$_{1g}$(Γ) | |
| | 580.6 ± 0.7 | 15 ± 4 | A$_{1g}$(M)+LA(M) + H$_2$Pc modes | |
| | 695 ± 1 | 52 ± 5 | 4LA(M) + H$_2$Pc modes | |

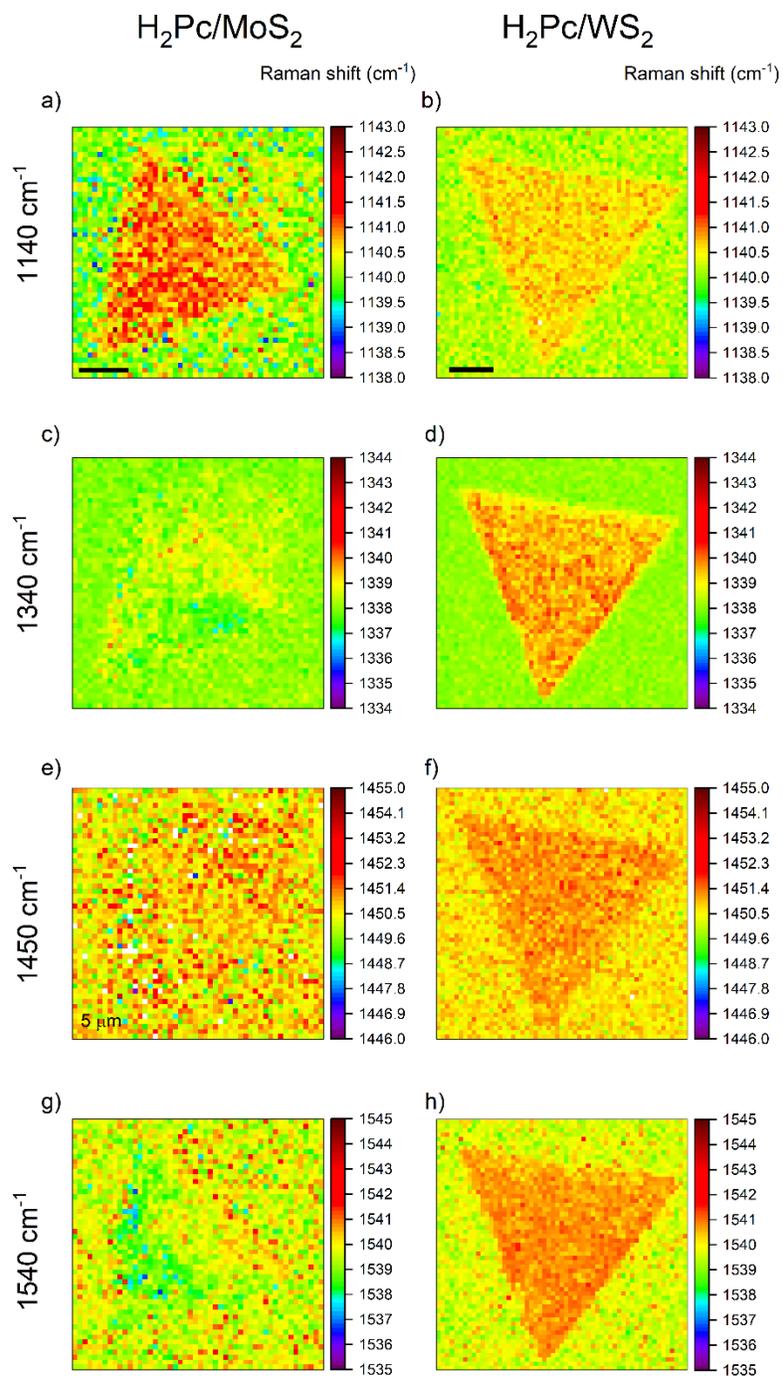

**Figure S5.** Spatial maps of H₂Pc Raman modes' frequencies for H$_2$Pc/MoS$_2$ (a,c,e,g) and H$_2$Pc/WS$_2$ (b,d,f,h) HSs. Scalebar is 5μm.